# Transitioning To The Digital Generation Case Studies (Previous Digital Point Studies In Japan Cases:1993-2023)

1st Yasuko Kawahata
*Rikkyo University*
*College of Sociology*
Department of Communication and Media Studies
ykawahata@rikkyo.ac.jp

*Abstract*—This Paper is Part of the 22nd IEEE/WIC International Conference on Web Intelligence and Intelligent Agent Technology (WI-IAT'23), The 8th International Workshop on Application of Big Data for Computational Social Science,(WI=ArtificialIntelligenceinthe Connected World October 26-29, 2023, Venice, Italy). To achieve the realization of the Global and Innovation Gateway for All (GIGA) initiative (2019), proposed in December 2019 by the Primary and Secondary Education Planning Division of the Elementary and Secondary Education Bureau of the Ministry of Education, Culture, Sports, Science and Technology, a movement has emerged to utilize information and communication technology (ICT) in the field of education. The history of ICT education in Japan dates back to the 100 Schools Project (1994), which aimed to provide network access environments, and the New 100 Schools Project (1997), which marked the beginning of full-scale ICT education in Japan. However, Japan continues to face a social issue, namely, the serious shortage of digital human resources; simultaneously, the new curriculum guidelines have started to make programming compulsory in elementary and junior high schools in order to enhance digital literacy. Accordingly, the GIGA initiative (2019)—a policy to distribute one tablet terminal to every elementary and junior high school student in Japan—and online classes have been rapidly promoted following the outbreak of the COVID-19 pandemic in 2020. In this paper, we discuss the usage dynamics of smartphone-based learning applications among young people (analyzing data from January–September 2020) and their current status. Further, the results are summarized and future research topics and issues are discussed. The results show that there are situations in which ICT learning environments can be effectively utilized and others in which they cannot, depending on the differences between digital students and analog students who utilize ICT in their studies; this indicates that we are currently in a transition to a generation of digital natives. ICT education has both advantages and disadvantages, and it is expected that it will be used in combination with conventional educational methods while assessing the characteristics of ICT education in the future.

Identify applicable funding agency here. If none, delete this.

## I. INTRODUCTION

In Japan, the term "information and communication technology (ICT) education" refers to the use of ICTs as an educational method. Accordingly, the concept of "everywhere computing" was formulated [1-8]. In response, the Advanced Information Technology Program was formulated in May 1994. This program included various principles as part of the informatization of education. Key among such principles is the realization of education and learning that transcend the limitations of classroom teaching. This goal is made possible by the use of computers and networks. Therefore, the Educational Software Development and Utilization Promotion Project was adopted as one of the Advanced Utilization of Specific Programs projects in the third supplementary budget for FY1993. One of the first major experiments involving ICT and elementary education in Japan was the 100 Schools Project, which began as part of the activities of the Network Operation Center, aiming to provide network access environments. This project focused on the procurement and installation of computer systems, software, modems, routers, and other equipment in target schools and the Network Operation Centers of the regional networks to which the target schools were connected; additionally, the telecommunication lines to be rented were all contracted by the IPA. From the end of FY1994 to early FY1995, the equipment was delivered and set up at the selected locations. In 1994, when the 100 Schools Project was launched, the above-mentioned academic and non-academic/non-profit regional networks were selected to connect approximately 100 target schools, which were distributed across all prefectures, to the Internet [33-36]. Subsequently, these regional networks were identified as connection points. Currently, ICT education can be broadly classified into the following five categories. As mentioned above, the 100 Schools Project (launched in 1994) was the beginning of full-fledged ICT education in Japan. Subsequently, the New 100 Schools Project was launched in 1997, covering 108 schools; this project emphasized diversity and also solicited research on children in various school settings, joint presentations, and other voluntary projects. Early on, initiatives such as the Educational Rating System Operation Experiment, the Utilization of Fixed Point Observation Data, and the Utilization of Existing Databases were undertaken as advanced projects, while themes such as international exchange, regional activities, and collaborative and integrated learning were also discussed, marking the beginning of intense activity. The movement to utilize ICT in the education field has been expanding toward the realization of the Global and Innovation Gateway for All (GIGA) initiative, proposed in December 2019 by the Primary and Secondary Education Planning Division of the Primary and Secondary Education Bureau of the Ministry of Education, Culture, Sports, Science and Technology (MEXT). However, in Japan, the shortage of digital human resources is becoming a serious social issue, while new curricula are starting to make programming compulsory in elementary and junior high schools in order to improve digital literacy. Moreover, the GIGA initiative, which aims to distribute one tablet terminal to every elementary and junior high school student in Japan, and online classes have been rapidly promoted following the outbreak of the COVID-19 pandemic in 2020 [9-20]. While these circumstances have made it possible to diversify learning and have been a step toward eliminating educational disparities, some scholars have questioned the effectiveness of the program in terms of its vision and impact on learning, leading to the emergence of two sides divided on this issue. As the Japanese educational system is rapidly changing, it is necessary to examine the effects and disadvantages of each approach and address them. Further, there is an urgent need to reduce the burden on teachers, who are at the frontline of the educational process. The purpose of this study is to summarize the past implementation status of ICT education in Japan, aiming to understand its future prospects and issues, as well as the related efforts in the industry. It also aims to conduct a questionnaire survey [2,3] to examine the usage dynamics of smartphone-based educational application (based on the Log Data Analysis of Smartphone Use (LDASU) project's data set, obtained from Fuller Corporation) as an example of effective ICT utilization in education. The results are summarized as a case study. Based on the discussion of the case study, we summarize future research issues regarding the role of ICTs in education to solve social problems in Japan, such as educational disparities and the lack of digital human resources, which can be complemented by the digital environment. Particularly, a major research question in this study concerns the usage and search dynamics of education-related applications listed on the Google Play store (which was analyzed based on the dataset), which play a part in ICT education. Subsequently, we summarize the current status of the usage dynamics (January–September 2020) of smartphone-based learning applications (Google Play) among young people engaged in learning.

## II. THE HISTORY OF ICT EDUCATION AND CASES OF ICTs' USE IN PRIVATE EDUCATION BEFORE COVID-19

### A. The 100 Schools Project and the New 100 Schools Project

ICT education in Japan has developed for several decades, facilitated by the widespread use of personal computers and the subsequent spread of wi-fi and the miniaturization of ICT devices [1-3]. This section summarizes the history of ICT education in Japan. The 100 Schools Project (which aims to establish network access environments across Japanese schools) and the New 100 Schools Project marked the beginning of full-scale ICT education in Japan. In order to develop and promote computer education in Japan, a plan was announced to introduce the Internet to all junior high schools, high schools, and schools for the disabled by 2001, as well as all elementary schools by 2003. The call for applications was distributed to the 47 prefectural boards of education nationwide in August 1994, under the guidance of the Ministry of Education [33-36]. Applications were divided into the following two groups: Group A consisted of approximately 20 to 30 schools that were particularly advanced and had the technical skills and real face of teachers; Group B consisted of approximately 70 to 80 schools that were able to actively plan and participate in network use and planning. Consequently, a total of 708 schools applied for Group A and 835 schools applied for Group B. Selection of eligible schools was based on a document-screening process in accordance with the Application Guidelines for the Network Usage Environment Provision Project. At least one eligible school was assigned to each prefecture; eligible schools included elementary schools, junior high schools, high schools, schools for the blind, schools for the deaf, and schools for the disabled, as well as international schools, in-hospital classes for children undergoing long-term treatment in hospitals, and three audio-visual centers where advanced efforts are expected. In June 1995, the connection to the Internet was completed and the servers became operational. The main examples of the utilization of the network include the dissemination of information using the web, presentation of learning results, collection and exchange of information and opinions, joint observation of various events in collaboration with specific partner schools, and international exchange with schools abroad. With the completion of the 100 Schools Project, the New 100 Schools Project was launched in 1997, covering 108 schools. Initially, a plan was announced to introduce the Internet into all junior high schools, high schools, and schools for the disabled by 2001, as well as all elementary schools by 2003, in order to develop computer education in Japan. The results of the 100 Schools Project and the New 100 Schools Project have encouraged the introduction of the Internet into the educational field.

## III. INITIATIVES IN THE PRIVATE SECTOR

The first case of ICT education in Japan in general was the "Satelline Seminar," in which Yoyogi Seminar started real-time delivery of lectures nationwide in 1989. After that, the cram school industry pioneered ICT education, as seen in the video classes offered by Tojin High School and Kawaijuku Manabis. This was very effective as a means of eliminating one of the educational disparities: the gap in the availability of education due to differences in residential areas. ICT education has various advantages in terms of cost, as it enables students to take classes at any time of the day, and reduces the total number of instructors. This educational method has attracted attention as a new form of learning, especially since many people have succeeded in passing the entrance examinations of the schools of their choice, and many have been accepted into good universities, including those that are considered difficult to enter. Further, opportunities for ICT education's utilization have increased. Correspondence courses are one of the private educational programs offered through online media. Benesse Corporation's Shinkenzemi (Challenge Touch), Z-kai of Z-kai Holdings, Inc. and Smile Zemi of Just System Inc. are particularly famous correspondence courses. Smile Zemi started its service in December 2012, and is characterized by the fact that it offers only one course that is completed entirely on a tablet device, without paper-based materials.

Challenge Touch was launched in April 2014 and had more than 1 million users as of 2017. Compared with the Shinkenzemi Elementary School Course, the subscription fee is similar, but students can choose either Challenge Touch, which is taught on tablet devices, or Challenge, which is taught using paper materials. The fees vary depending on the course and grade. However, details on the number of users of Shinkenzemi (Challenge Touch), Zukai, and Smile Zemi are not disclosed. The Classroom of the Future demonstration project being promoted by the Ministry of Economy, Trade and Industry since FY2018 aims to share examples of learning and services utilizing Edtech. In addition to initiatives at schools across Japan, many educational services provided by private companies have been adopted. One such service is See-be, by Sanaru Corporation, a company in the cram school industry, which has been installed in all of Sanaru Prep School's buildings and classrooms. See-be is a tool that enables accurate and easy-to-understand animations and historical materials based on a vast amount of data, as well as simulations of experiments operated by the teacher, to be projected on the whiteboard. Moreover, Kumon's ICT education is being gradually introduced into other fields besides tutoring schools that offer video lessons. Uchida Yoko, which was originally an office-equipment trading company, is now involved in designing learning spaces and developing and selling products related to ICT education, in addition to its existing business. In addition, Uchida Yoko holds an annual event called New Education Expo as a place to share information on the education field.

*A. Recent ICT education initiatives for primary education in Japan*

The Five-Year Plan for Environmental Improvement for ICT in Education (FY 2018-2022), formulated in April 2018, compiled policies for improving the ICT environment based on the contents of the new Courses of Study (i.e., curricula) to be revised in FY 2020-2022 for elementary through high schools. Specifically, these new study guidelines for elementary schools include mandatory programming education. The Policy for the Improvement of ICT Environment in Schools from FY 2018 Onward set the following seven goals for the improvement of the ICT environment in schools, requiring that 180.5 billion yen be allocated for a single fiscal year (FY 2018 to FY 2022). (1) Computers for learners: Provide one computer for every three classes. (2) Computers for instructors: one for each teacher in charge of a class (3) Large-screen presentation devices and actual projectors: 100Actual projectors will be installed in elementary schools and special-needs schools based on the actual maintenance status. (4) 100Additionally, various servers and security software are also covered. According to the 2018 PISA ICT utilization survey, Japan ranked last among Organisation for Economic Co-operation and Development-member countries, measured as the time spent using digital devices in class. The Five-Year Environmental Improvement Plan for ICT in Education (FY 2018-2022) is the first step in a policy to begin focusing on ICT education within this context, where ICT education is not widespread.

*B. Revision of the law on digital textbooks for learners*

There are two types of digital textbooks: "instructor textbooks," which are used by teachers in class, and "learner textbooks," which are used by students on their own computers, tablets, etc., in the same way as paper textbooks. While digital textbooks for instructors have been introduced and are being used in the classroom, digital textbooks' introduction has lagged behind. This is due to the fact that digital textbooks for learners were not recognized as textbooks, while copyright fees, which are typically waived for paper textbooks, had not been waived for digital textbooks. Therefore, on June 1, 2018, the MEXT promulgated the Law Partially Amending the School Education Law, which allows learner digital textbooks containing the same content as paper textbooks to be used in place of and in conjunction with paper textbooks, starting on April 1, 2019. Thus, since digital textbooks for learners are now treated as textbooks under this law, copyright fees for digital textbooks for learners are now handled in the same way as for paper textbooks, in accordance with Article 33 of the Copyright Law. However, restrictions remain on the use of digital textbooks due to the lack of computers and tablet terminals for use by students and the fact that digital textbooks could only be used for less than one-half of the number of class hours for each subject under Article 34, Paragraph 2 of the School Education Law. Additionally, while paper textbooks were provided free of charge to students at government expense, digital textbooks were not, and the cost of 200 to 2,000 yen per subject was borne by the board of education. Therefore, the introduction of digital textbooks has not progressed adequately. According to a survey by the MEXT, digital textbooks for elementary school students were available in 20The Law Partially Amending the School Education Law, which went into effect in FY 2019, restricted digital textbooks' usage time, out of consideration for children's health, such as eye fatigue; however, this was criticized by the expert committee members as having no basis, while considering the spread of COVID-19 and the GIGA initiative. On December 22, 2020, a proposal to eliminate the standard limiting the usage time for digital textbooks to "less than one-half the number of class hours for each subject" was approved, and digital textbooks for learners became available for use in place of paper textbooks, starting on April 1, 2021. Regarding health concerns, the MEXT pointed out that "regardless of whether it is paper or digital, prolonged and continuous close gazing should be avoided from the perspective of vision deterioration," and requested that when using a terminal in class, students (1) take their eyes off the screen for about 20 seconds every 30 minutes and rest, (2) maintain a good posture and keep a distance of 30 to 50 cm between their eyes and the terminal, among other precautions. The MEXT is currently working on the GSE and considering the full-scale introduction of digital textbooks for learners to be used with the terminals developed under the GIGA initiative in FY2024, when the next revision of elementary school textbooks is scheduled. The government plans to raise the rate to 100

*C. The GIGA initiative and private sector initiatives in the COVID-19 vortex*

The GIGA initiative, launched by the MEXT in December 2019, is a policy to promote ICT education in elementary and junior high schools. According to the MEXT website, "By integrating one terminal per student and a high-speed, large-capacity communication network, we will realize an educational environment where diverse children, including those with special needs, receive optimized education fairly and individually without leaving anyone behind, and where their qualities and abilities can be further cultivated without fail. The best mix of Japan's existing educational practices and state-of-the-art technology will be used to maximize the abilities of teachers and students." The program mainly includes subsidies for the cost of tablet terminals to be distributed to each student, as well as subsidies for the cost of developing a wi-fi environment. In the case of private elementary and junior high schools, the subsidy is half the cost of the device, with an upper limit of 45,000 yen per device. Originally, the GIGA initiative aimed to develop this hardware environment over a five-year period, starting in FY2019; however, the need for online classes increased due to the closure of schools due to the COVID-19 pandemic, which led to its accelerated implementation. According to the Nikkei Shimbun, in distributing one tablet terminal per student, only one terminal per 6.1 students in elementary schools and one per 5.2 students in junior high schools had been deployed by 2019, before the GIGA initiative went into full swing. In July 2020, a total of 74 municipalities in the 23 wards of Tokyo, prefectural cities, and government ordinance cities were examined. When asked about the status of securing terminals for public elementary and junior high schools, only Shibuya Ward, in Tokyo, had already deployed them; two municipalities, Nara City and Toshima Ward, in Tokyo, were able to complete deployment by September; and nine municipalities, including Sakai City, indicated they would be able to complete deployment from October to December. Of the 62 municipalities (83

At the New Education Expo 2020, which took place in October 2020, a wide range of events were held, including a hands-on experience with Future Classroom, a product of Uchida Yoko, which comprised a seminar on the pros and cons of digital textbooks for learners, as well as a seminar on the creation of universities that continue to be chosen by the public. The results of ICT utilization in private education, which has been promoted ahead of public education, may have had some influence on the efforts and policies to realize ICT utilization in public education. Furthermore, as an example of original efforts in public education, some schools had already begun to use electronic blackboards and tablet terminals for education before the GIGA initiative was launched. At Dai-Ichi Gakuin High School, a nationwide correspondence high school, all students had and were using tablets by 2015. Further, N High School, a correspondence high school that has been attracting attention in recent years for its rapidly growing enrollment rates, has been providing education as an "online high school," with all students owning tablet devices since its opening in 2016. In addition to using the chat tool Slack for online homerooms and communicating class schedules to homeroom teachers, VR equipment is provided free of charge to students in the regular course at N High School, allowing them to use VR learning software to view original class materials (as of March 2022, my writing)[37-40].

*D. The national and international contexts and other initiatives*

Despite these ongoing efforts, Japan's usage time for digital devices in the classroom ranked last among Organisation for Economic Co-operation and Development-member countries in the 2018 PISA survey on the use of ICTs. Countries ranked higher include Sweden, Denmark, Australia, and New Zealand. Recently, it has been said that Japan's ICT use is also inferior to that of South Korea and Singapore. For example, in Denmark, some students have been tested to bring in tablet devices to foster the ability to efficiently and accurately gather information on the Internet, while the government of Queensland, Australia, has created and released digital textbooks that can be used free of charge. Science, technology, engineering and mathematics education began in the U.S. in the 2000s and was made famous by former US President Barack Obama. The school, a pioneer in problem-based learning, focuses especially on programming and non-cognitive skills development, and actively adopts e-learning. With no tuition, no textbooks, and no grade book, students can develop the ability to think for themselves and realize their own goals while working in teams on projects. Additionally, the potential of ICT for utilization in special-needs education has been recognized for some time, as shown by the selection of schools for the blind, the deaf, and schools for the disabled in the aforementioned 100 Schools Project since 1994. The concept of universal design (UD) has permeated the design of electronic devices. For example, the iPhone's accessibility features include voice reading, magnification, black and white inversion, grayscale, UD fonts, and contrast enhancement, which can be customized according to the user's needs. In 2016, the UD Digital Textbook Style was released, featuring a simple shape with a constant thickness, developed with low-vision and visually-sensitive children in mind. For example, in the case of a junior high school student with dyslexia, the read-aloud function of a digital textbook is used. The printouts that have not been digitized can also be read out loud by using an application that can digitize them. For writing, the students look up the dictionary application for writing practice, write while looking at the shapes of the characters, and use an application that judges the neatness of the written characters. In the case of an elementary school student with high-functioning autism and LD, tablet devices are used for notetaking via keyboard input, sharing worksheets created by the teacher using PowerPoint, and remote management of learning status. The Act on the Elimination of Discrimination against Persons with Disabilities, enacted in 2016, has made the concept of reasonable accommodation more prevalent, and an increasing number of students are being allowed to take regular examinations and take examinations by computer or by voice substitution. However, there are also challenges in terms of the conflict owed to individual differences and the higher level required. Particularly, children with developmental disabilities tend to be more dependent on the Internet and video games, and there are physical challenges such as crimes and charges through social networking services (SNSs), back and eye problems, and sleep issues. LITALICO, a private company that provides rehabilitation and education for children with developmental disabilities, is following up on cases that are beyond the reach of the government, while the LITALICO Research Institute is working to improve the company's services to guarantee information, and other mechanisms to achieve accessibility and reasonable accommodation [41-43].

## IV. Prospects of this analysis

This paper examines the diffusion of ICT education in Japan and how it can be effectively used to solve social problems. In order to investigate policy and private education initiatives, as well as to clarify familiar personal use of ICT education, we analyzed the use of educational apps among junior and senior high school students. The analysis of LDASU data showed that the frequency of educational app use tended to increase among second- and third-year high school students during the college entrance exam period and during school commuting hours. Additionally, the analysis of the usage dynamics of other apps among users who frequently used educational apps showed that SNSs and news apps were used frequently, whereas game apps were not. This indicates that we are currently witnessing a transition to a digital native generation, and that there are situations in which ICTs can (or cannot) be effectively utilized in the learning of elementary, junior high, and high school students, with some students making use of ICT and others being analogists. It is assumed that ICT education will be used in combination with traditional educational methods, while carefully considering the characteristics of ICT education. While this paper found that students themselves tend to actively use learning applications in educational settings, it also highlighted the need to actively and safely utilize ICTs while using both ICT education and conventional educational methods as necessary. ICT education is expected to be a new form of education; however, at this stage, the main question is whether or not it can be used safely by children. It is clear that independent learning, which is emphasized in the new Courses of Study, can be realized only when its flexible use is allowed, and the benefits of ICT education's ability to provide individualized education can also be utilized. However, as in the case of NETS in the US, even if rules for use are established at each school, they may not function without a legal binding force. It can be said that the system and legislation for the safe use of ICTs—including its mobile (e.g., smartphone) and security aspects—in education must be rigorous. In terms of future prospects, it is desirable to verify what kind of results will be achieved when the current elementary and junior high school students, for whom programming has become compulsory, enter the workforce, and to use this as a reference for revising policies and curriculum guidelines in order to improve logical thinking skills and solve the shortage of digital human resources, which are expected to become a focal point of discussion when ICT education becomes widespread. As a point of focus for future analysis, it was also confirmed that middle and high school users who frequently used educational applications also frequently used communication applications such as SNSs (text- or image-centered platforms such as Twitter and Instagram) and blogs (Ameba). In terms of learning effectiveness in the digital environment, it can be assumed that the student generation also views the opinions of faculty and staff in the discourse space of SNSs, which can be easily accessed from their smartphones. In light of this, it will be necessary to analyze data from 2020 onward, especially within the context of the COVID-19 pandemic and the GIGA initiative, to see how elementary, middle, and high school teachers are increasingly intervening in the discourse space of SNSs.

## V. Acknowledgment

In conducting this research, we would like to express our sincere gratitude to all members of the Kawahata Seminar, Department of Sociology, Rikkyo University, for their discussion and research participation in this project from 2020 onward. We would like to express our sincere gratitude to T.M., a graduate of the Kawahata Seminar in the class of 2021, who worked with the author on the research, survey, and writing activities that form the basis of this paper. We would also like to express our sincere gratitude to Professor Shinichiro Wada and Professor Tadamasa Kimura of the Department of Media and Society, Faculty of Sociology, for providing the data set, UserLocal, Inc. and everyone who willingly cooperated in the interview and questionnaire surveys. We would also like to thank Fuller, Inc. for the LDASU2020 dataset and for their active research input and sharing of challenges during this study. We would also like to thank the JSPS/JSPS Grants-in-Aid for Scientific Research (JSPS/Grants-in-Aid for Scientific Research) for the fiscal years 2018-2021.Grant-in-Aid for Scientific Research Project Research Subject/Area No. 19K04881, "Construction of a New Theory of Opinion Dynamics that Can Describe the Real Image of Society by Introducing Trust and Distrust". Finally, we would like to express our sincere gratitude to Ishii Laboratory, Graduate School of Technology, Tottori University (until FY2022), who was our joint research team in the above Grant-in-Aid research project.